\magnification=\magstep1
\baselineskip=16pt                      
\font\title = cmr10 scaled 1440    
\newcount\ftnumber
\def\ft#1{\global\advance\ftnumber by 1
          {\baselineskip 12pt    
           \footnote{$^{\the\ftnumber}$}{#1}}}

  \def\ftn#1#2{{\baselineskip12pt\footnote{$^{#1}$}{#2}}}

\newcount\eqnumber

\def\sa#1#2{\sigma^#1_#2}

\def\sc#1#2#3{\sigma^1_#1\sigma^2_#2\sigma^3_#3}
\def\sx{\sigma_x}
\def\sy{\sigma_y}
\def\nv{{\bf n}}
\def\mv{{\bf m}}
\def\av{{\bf a}}
\def\bv{{\bf b}}
\def\cv{{\bf c}}
\def\dv{{\bf d}}
\def\ev{{\bf e}}
\def\fv{{\bf f}}
\def\gv{{\bf g}}
\def\xv{{\bf x}}
\def\yv{{\bf y}}
\def\zv{{\bf z}}

\def\;{\,\,}

\def\<{\langle}
\def\>{\rangle}
\newcount\eqnumber
\def\equ(#1){\global\advance\eqnumber by 1 
    \expandafter\xdef\csname !#1\endcsname{\the\eqnumber}
    \eqno(\the\eqnumber)}                               
\def\(#1){(\csname !#1\endcsname)}
\def\fr#1/#2{{\textstyle{#1\over#2}}} 

\def\boxit#1{\vbox{\hrule\hbox{\vrule\kern3pt         
     \vbox{\kern3pt#1\kern3pt}\kern3pt\vrule}\hrule}} 
\def\ie{{\it i.e.\/}$\,\,\,$} 

%
\newbox\FatInternalVariable		
\def\fat#1{{
\hbox{					
\setbox\FatInternalVariable=\hbox{#1}
\unhcopy\FatInternalVariable
\kern-\wd\FatInternalVariable
\kern 0.015 em
\unhcopy\FatInternalVariable
\kern-\wd\FatInternalVariable
\kern 0.030 em
\unhbox\FatInternalVariable
}
}}
\def\nhv{no-hidden-variables}
\def\hvt{hidden-variables theory}
\def\bks{Bell-KS}
\def\bt{Bell's Theorem}
\def\ghz{Greenberger, Horne, and Zeilinger}
\def\head#1{\leftline{\bf #1}}

\centerline{\title Hidden Variables and the Two Theorems of John Bell}
\vskip 20pt
\centerline{N. David Mermin}
\vskip 20pt

What follows is the text of my article 
that appeared in Reviews of Modern Physics {\bf 65}, 803-815 (1993). 
 I've corrected the three small errors posted over the years at the Revs.~Mod.~Phys.~website.    I've removed two decorative figures and changed the other two figures into (unnumbered) displayed equations.   I've made a small number of minor editorial improvements.    I've added citations to some recent critical articles by Jeffrey Bub and Dennis Dieks.   And I've added a few footnotes of commentary.\ftn*{The added footnotes are denoted by single or double asterisks.  The footnotes from the original article have the same numbering as in that article.} 
 \bigskip
 
I'm posting my old paper at arXiv for several reasons.   First, because it's still timely and I'd like to make it available to a wider audience in its 25th anniversary year.  
Second, because, rereading it, I was struck that Section VII raises some questions that, as far as I know, have yet to be adequately answered.   And third, because it has recently been criticized by Bub and Dieks as part of their broader criticism of John Bell's and Grete Hermann's reading of John von Neumann.\ftn{**}{R\"udiger Schack and I will soon post a paper  in which we give our own view of von Neumann's four assumptions about quantum mechanics, how he uses them to prove his famous no-hidden-variable theorem, and how he fails to point out that one of his assumptions can be violated by a hidden-variables theory without necessarily doing violence to the whole structure of quantum mechanics.  Schack and I explain how Bub and Dieks miss this point in their criticisms of Bell and Hermann.    Bell and Hermann have indeed correctly identified von Neumann's oversight, as I described, perhaps too briefly,  in the review article reproduced below.}

\vfil\eject
\centerline{{\title Hidden Variables and the Two Theorems of John Bell}
}
\medskip
\bigskip
\centerline{N. David Mermin}
\centerline{Laboratory of Atomic and Solid State Physics}
\centerline{Cornell University, Ithaca, NY 14853-2501}
\bigskip

\midinsert
\narrower\narrower 

{\it Abstract.\/} Although skeptical of the prohibitive power of
no--hid\-den--variables theorems, John Bell was himself responsible for the
two most important ones.  I describe some recent versions of the lesser
known of the two (familar to experts as the ``Kochen--Specker theorem'')
which have transparently simple proofs.  One of the new versions can be
converted without additional analysis into a powerful form of the very much
better known ``Bell's Theorem'', thereby clarifying the conceptual link
between these two results of Bell.  

\endinsert

\bigskip

\midinsert
\narrower
\noindent
{\it Like all authors of noncommissioned reviews he thinks that he
can restate the position with such clarity and simplicity that all previous
discussions will be eclipsed.} \null\hskip4.5truein --- J. S. Bell, 1966
\endinsert
\bigskip
\noindent CONTENTS
\medskip

\item{I.}The Dream of Hidden Variables
\item{II.} Plausible Constraints on a hidden-variables Theory
\item{III.} Von Neumann's Silly Assumption
\item{IV.} The Bell--Kochen--Specker Theorem
\item{V.} A Simpler Bell-KS Theorem in 4 Dimensions
\item{VI.} A Simple and More Versatile Bell-KS Theorem in 8 Dimensions
\item{VII.}  Is the Bell-KS Theorem Silly?
\item{VIII.}  Locality Replaces non-contextuality: Bell's Theorem 
\item{IX.}  A Little about Bohm Theory
\item{X.} The Last Word
\item{}Acknowledgments
\item{}References

\bigskip
\head{I. THE DREAM OF HIDDEN VARIABLES}
\medskip

It is a fundamental quantum doctrine that a measurement does not, in
general, reveal a pre-existing value of the measured property.  On the
contrary, the outcome of a measurement is brought into being by the act of
measurement itself, a joint manifestation of the state of the probed system
and the probing apparatus.  Precisely how the particular result of an
individual measurement is brought into being---Heisenberg's ``transition
from the possible to the actual''---is inherently unknowable.  Only the
statistical distribution of many such encounters is a proper matter for
scientific inquiry.

We have been told this so often that the eyes glaze over at the words, and
half of you have probably stopped reading already.  But is it really true?
Or, more conservatively, is it really necessary?  Does quantum mechanics,
that powerful, practical, phenomenally accurate computational tool of
physicist, chemist, biologist, and engineer, really demand this weak link
between our knowledge and the objects of that knowledge?  Setting aside the
metaphysics that emerged from urgent debates and long walks in Copenhagen
parks, can one point to anything in the modern quantum theory that forces on
us such an act of intellectual renunciation?  Or is it merely reverence for
the Patriarchs that leads us to deny that a measurement reveals a value that
was already there, prior to the measurement?

Well, you might say, it's easy enough to deduce from quantum mechanics that
in general the measurement apparatus disturbs the system on which it acts.
True, but so what?  One can easily imagine a measurement messing up any
number of things, while still revealing the value of a pre-existing
property.  Ah, you might add, but the uncertainty principle prohibits the
existence of joint values for certain important groups of physical
properties.  So taught the Patriarchs, but as deduced from within the
quantum theory itself, the uncertainty principle only prohibits the
possibility of preparing an ensemble of systems in which all those
properties are sharply defined; like most of quantum mechanics, it
scrupulously avoids making any statements whatever about individual members
of that ensemble.  But surely indeterminism, you might conclude, is built
into the very bones of the modern quantum theory.  Entirely beside the
point!  The question is whether properties of individual systems possess
values prior to the measurement that reveals them; not whether there are
laws enabling us to predict at an earlier time what those values will be.

What, in fact, can you say if called upon to refute a celebrated polymath
who confidently declares\ftn{*}{Adler, 1992, p. 300.} that ``Most theoretical
physicists are guilty of$\ldots$fail[ing] to distinguish between a
measurable indeterminacy and the epistemic indeterminability of what is in
reality determinate.  The indeterminacy discovered by physical measurements
of subatomic phenomena simply tells us that we cannot know the definite
position and velocity of an electron at one instant of time.  It does not
tell us that the electron, at any instant of time, does not have a definite
position and velocity.  [Physicists]$\ldots$convert what is not measurable
by them into the unreal and the nonexistent.'' 

Are we, then, arrogant and irrational in refusing to consider the
possibility of an expanded description of the world, in which properties
such as position and velocity do have simultaneous values, even though
nature has conspired to prevent us from ascertaining them both at the same
time?  Efforts to construct such deeper levels of description, in which
properties of individual systems do have pre-existing values revealed by the
act of measurement, are known as hidden-variables programs.  A frequently
offered analogy is that a successful hidden-variables theory would be to
quantum mechanics as classical mechanics is to classical statistical
mechanics:\ftn{*}{See, for example, A.  Einstein in Schilpp, 1949, p. 672.}
quantum mechanics would survive intact, but would be understood in terms of
a deeper and more detailed picture of the world.  Efforts, on the other
hand, to put our notorious refusal on a more solid foundation by
demonstrating that a hidden-variables program necessarily requires outcomes
for certain experiments that disagree with the data predicted by the quantum
theory, are called no-hidden-variables theorems (or, vulgarly, ``no-go
theorems'').  

In the absence of any detailed features of a hidden-variables program,
quantum mechanics is incapable of demonstrating that the general dream is
impossible.\ft{David Bohm (Bohm, 1952) has, in fact, provided a
hidden-variables theory that, if nothing else, serves as a proof that an
unqualified refutation is impossible.  I will return to Bohm theory in
Section IX, merely noting here that it does exactly what Mortimer Adler
wants, while remaining in complete agreement with quantum mechanics in its
predictions for the outcome of any experiment.} If the program consists of
nothing beyond the bald assertion that such values exist, then while quantum
physicists may protest, the quantum theory is powerless to produce a case
where experimental data can refute that claim, precisely because the theory
is mute on what goes on in individual systems.  A hidden-variables theory
has to make {\it some\/} assumptions about the character of those
pre-existing values if quantum theory is to have anything to attack.

John Bell proved two great no-hidden-variables theorems.  The first, given in
Bell, 1966, is not as well known to physicists as it is to philosophers, who
call it the Kochen-Specker (or KS) theorem\ft{As mathematics, both results
are special cases of a more powerful analysis by A. M.  Gleason, 1957.}
because of a version of the same argument, apparently more to their taste,
derived independently by S.  Kochen and E. P.  Specker, 1967.  I shall refer
to it as the Bell-KS theorem.  The second theorem, ``Bell's Theorem'', is
given in Bell, 1964,\ft{In spite of the earlier publication date Bell's
Theorem was proved {\it after\/} Bell proved his 1966 theorem.  The
manuscript of Bell, 1966, languished unattended for over a year in a drawer
in the editorial offices of {\it Reviews of Modern Physics\/}.} and is
widely known not only among physicists, but also to philosophers,
journalists, mystics, novelists, and poets.

One reason the \bks\ theorem is the less celebrated of the two is that the
assumptions made by the hidden-variables theories it prohibits can only be
formulated within the formal structure of quantum mechanics.  One cannot
describe the Bell-KS theorem to a general audience, in terms of a
collection of black-box {\it gedanken\/} experiments, the only role of
quantum mechanics being to provide {\it gedanken\/} results, which all by
themselves imply that at least one of those experiments could not have been
revealing a pre-existing outcome.  Bell's Theorem, however, can be cast in
precisely such terms.\ft{Several such formulations of Bell's Theorem are
given in Mermin, 1990a.} Indeed the hidden-variables theories ruled out by
\bt\ rest on assumptions that not only can be stated in entirely
non-technical terms but are so compelling that the establishment of their
falsity has been called, not frivolously, ``the most profound discovery of
science''.\ftn{*}{Stapp, 1977.} 

The comparative obscurity of the Bell-KS theorem may also derive in part
from the fact that the assumptions on which it rests were severly and
immediately criticized by Bell himself: ``That so much follows from such
apparently innocent assumptions leads us to question their innocence''.  We
shall return to his criticism in Section VII.

A less edifying reason for the greater fame of \bt\ among physicists is that
its proof is utterly transparent, while proving the Bell-KS theorem entails
a moderately elaborate exercise in geometry.  Physicists are simply less
willing than philosophers to suffer through a few pages of dreary analysis
to prove something they never doubted in the first place.  So although all
physicists know about \bt, most look blank when you mention Kochen-Specker
or \bks.  Now, however, these particular grounds for such ignorance have been
removed.  Within the past few years new versions of the
\bks\ theorem have been found\ftn{**}{Mermin, 1990b.} that are so simple that even
those physicists who regard such efforts as pointless, can grasp the
argument with negligible waste of time and mental energy.  Besides making
the argument so easy that even impatient physicists can enjoy it, one of the
new forms of the \bks\ theorem can also be readily converted into the
striking new version of Bell's Theorem invented by \ghz,\ft{Greenberger,
1989.  I have given a concise version of the Greenberger-Horne-Zeilinger
argument in Mermin, 1990c and 1990d.  An expanded discussion of their
original argument can be found in Greenberger, 1990.} thereby shedding a new
light on the relation between these two results of Bell.

\bigskip
\head{II. PLAUSIBLE CONSTRAINTS ON A HIDDEN-VARIABLES THEORY} 
\medskip

I now specify more precisely the general features of a hidden-variables
theory.  Quantum mechanics deals with a set of observables $A, B, C,\ldots$
and a set of states $|\Psi\>$, $|\Phi\>,\ldots\,$.  If we are given a
physical system described by a particular state, then quantum mechanics
gives us the probability of getting a given result when measuring one of the
observables.  More generally, if we have a group of mutually commuting
observables, quantum mechanics asserts that we can do an experiment that
measures them simultaneously, and gives us the joint distribution for the
values of each of the observables in that mutually commuting set.

We wish to entertain the heretical view that the results of a measurement
are not brought into being by the act of measurement itself.  This heresy
takes the state vector to describe an ensemble of systems, and maintains
that in each individual member of that ensemble every observable does indeed
have a definite value which the measurement merely reveals when carried out
on that particular individual system.  The quantum mechanical rules, applied
to a given state, give the statistics obeyed by those definite values in the
ensemble described by that state. The uncertainty principle is not a
restriction on the ability of observables to possess values in individual
systems, but a limitation on the kinds of ensembles of individual systems it
is possible to prepare, stemming from the unavoidable disturbance the
state-preparation procedure imposes on the system.  If two observables fail
to commute then the uncertainty principle does not prohibit both from having
definite values in an individual system.  It merely insists that it is
impossible to prepare an ensemble of systems in which the values of neither
observable fluctuate from one individual system to another.

To this kind of talk the well-trained quantum mechanician says ``Rubbish!''
and gets back to serious business.  But is it possible to offer a better
rejoinder?  Is it possible to demonstrate not only that the innocent view is
at odds with the prevailing orthodoxy, but that it is, in fact, directly
refuted by the quantum mechanical formalism itself, without any appeal to an
interpretation of that formalism?  A \nhv\ theorem attempts to provide such
a refutation.  It is only an attempt because any such theorem must make some
assumption on the nature of the hidden variables it excludes, which a
persistent heretic can always call into question.  Here is what I hope you
will agree are a plausible set of assumptions for a straightforward
hidden-variables theory:\ft{But in VII we will come back, with Bell, to
criticize one of them, so look them over carefully!  At this point I
deliberately refrain from calling the elusive culprit to your attention.  It
is my hope that you will finding the assumptions sufficiently harmless to be
curious whether any hidden-variables theory meeting such apparently benign
conditions can indeed be ruled out by hard-headed quantum-mechanical
calculation, rather than merely being rejected because it is in bad taste.}

Given an ensemble of identical physical systems all prepared in the state
$|\Phi\>$ described by observables $A, B, C,\ldots$ such a theory should
assign to each individual member of that ensemble a set of numerical values
for each observable, $v(A), v(B), v(C)\,\ldots$, so that if any observable or
mutually commuting subset of observables is measured on that individual
system, the results of the measurement will be the corresponding
values.\ft{Whether, and in what way, those values depend on new parameters
or degrees of freedom is a detail of the particular hidden-variables theory,
and plays no role in what follows, except for the 2-dimensional example of
Bell described below.} The theory should provide a rule for every state
$|\Phi\>$ telling us how to distribute those values over the members of the
ensemble described by $|\Phi\>$ in such a way that the statistical
distribution of outcomes for any measurement quantum mechanics permits, agree
with the predictions of quantum mechanics.

Some of the constraints quantum mechanics imposes on the values are
independent of the state $|\Phi\>$ we are examining.  In particular quantum
mechanics requires that the result of measuring an observable must be an
eigenvalue of the corresponding hermitian operator.  Therefore only the
eigenvalues of $A$ can be allowed as values $v(A)$.  Quantum mechanics
further requires that if $A, B, C,\ldots$ are a mutually commuting subset of
the observables then the only allowed results of a simultaneous measurement
of $A, B, C,\ldots$ are a set of simultaneous eigenvalues.  This
correspondingly restricts the set of values $v(A), v(B), v(C), \ldots$
possessed by any individual system.  In particular, since any functional
identity $$f(A,B,C,\ldots) = 0\equ(fA)$$ satisfied by a mutually commuting
set of observables is also satisfied by their simultaneous eigenvalues it
follows that if a set of mutually commuting observables satisfies a relation
of the form \(fA) then the values assigned to them in an individual system
must also be related by $$f(v(A),v(B),v(C),\ldots) = 0.\equ(fvA)$$ 

Remarkably, some \nhv\ theorems arrive at a counterexample by considering
only \(fA) and
\(fvA), without even needing to appeal to the further constraints on the
values impossed by the statistical properties of a particular state.  The
Bell-KS theorem is such a result.  Others, of which Bell's Theorem is the
most important example, require the properties of a special state to
construct counterexamples.  We shall examine in VIII why it might be
necessary for the scope of the counterexample to be restricted in this way.
But before we begin, let us first look at a famous false start.

\bigskip
\head{III. VON NEUMANN'S SILLY ASSUMPTION}
\medskip

Many generations of graduate students who might have been tempted to try to
construct hidden-variables theories were beaten into submission by the claim
that von Neumann, 1932, had proved that it could not be done.  A few years
later\ftn*{See Jammer, 1974, p.~273.} Grete Hermann, 1935, pointed out a
glaring deficiency in the argument, but she seems to have been entirely
ignored.  Everybody continued to cite the von Neumann proof.  A third of a
century passed before John Bell, 1966, rediscovered the fact that von
Neumann's \nhv\ proof was based on an assumption that can only be described
as silly\ft{While giving a physics colloquium on these matters I was taken
to task by an outraged member of the audience for using the adjective
``silly'' to characterize von Neumann's assumption.  I subsequently
discovered that, like many penetrating observations about quantum mechanics,
this one was made emphatically by John Bell: ``Yet the von Neumann proof, if
you actually come to grips with it, falls apart in your hands!  There is {\it
nothing\/} to it.  It's not just flawed, it's {\it silly\/}!$\,\ldots$ When you
translate [his assumptions] into terms of physical disposition, they're
nonsense.  You may quote me on that:  The proof of von Neumann is not merely
false but {\it foolish\/}!'' (Interview in {\it Omni\/}, May, 1988, p.
88.)}---so silly,\ftn{**}{Recently Bub, 2010, 2011, and Dieks, 2017, have argued that Bell and Hermann had misunderstood von Neumann.   I disagree.  R\"udiger Schack and I are writing a commentary on von Neumann's argument, that contains our criticism of Bub and Dieks.}  
in fact that one is led to wonder whether the proof was
ever studied by either the students or those who appealed to it to rescue
them from speculative adventures.  

A particular consequence of
\(fA) and \(fvA) is that if $A$ and $B$ commute then the value assigned to
$C = A + B$ must satisfy $$v(C) = v(A) + v(B),\equ(vsum)$$ as an expression
of the identity $C - A - B = 0$.  Von Neumann's silly assumption was to
impose the condition \(vsum) on a hidden-variables theory even when $A$ and
$B$ do not commute.  But when $A$ and $B$ do not commute they do not have
simultaneous eigenvalues, they cannot be simultaneously measured, and there
are absolutely no grounds for imposing such a requirement.  Von Neumann was
led to it because it holds in the mean:  for any state $|\Phi\>$, quantum
mechanics requires, whether or not $A$ and $B$ commute, that
$$\<\Phi|A+B|\Phi\> =
\<\Phi|A|\Phi\> +
\<\Phi|B|\Phi\>.\equ(mean)$$ But to require that $v(A+B) = v(A) + v(B)$ in
each individual system of the ensemble is to ensure that a relation
holds in the mean by imposing it case by case---a sufficient, but hardly a
necessary condition.  Silly!\ftn*{Rereading these last two sentences in the light of the defense of von Neumann offered by Bub, 2010, 2011,  and Dieks, 2017, I realize that I am giving too abbreviated a summary of what von Neumann actually says.   Could it be just this cartoon of his argument that is silly?   In my forthcoming paper with Schack we  demonstrate that  in his full, carefully articulated development von Neumann actually does commit a genuine blunder, clearly identified by Bell,  the best one-word characterization of which I continue to believe is ``silly''.}

That the results of quantum mechanics are incompatible with values
satisfying this condition is easy to see even in the two-dimensional state
space that describes a single spin--$\fr1/2$.  Let $A=\sx,\ B=\sy$. The
eigenvalues of the Pauli matrices are $\pm 1$ so the values $v(A)$ and
$v(B)$ are each restricted to be $\pm 1$. Thus the only values $v(A)+v(B)$
can have are $-2, 0,$ and 2.  But $A+B$ is just $\sqrt2$ times the component
of\fat{$\sigma$}along the direction bisecting the angle between the $x$-- and
$y$--axes.  As a result its allowed values are $v(A+B) = \pm\sqrt2$.
Therefore a hidden-variables theory of this simple system cannot satisfy
\(vsum).  But there is no reason to insist that it should!  Indeed, having
exposed the silliness in the von Neumann argument, Bell went immediately on
to construct a hidden-variables model for a single spin--$\fr1/2$ that
satisfies all the non-silly conditions specified above.  I now give this
construction, but include it only to emphasize the non--triviality of the
impossibility proofs we shall then turn to.  Readers not interested in the
details of Bell's counterexample can skip to Section IV.

In a two-dimensional state space every state is an eigenstate of the
component $\sigma_n$ of the spin along some direction \nv:\ft{This is
because every state can be related to $|\uparrow_z\>$ by a unitary
transformation, but in a 2-dimensional state space any unitary
transformation, being a member of SU(2), represents a rotation.} $$
\sigma_n|\uparrow_n\> = |\uparrow_n\>,\equ(sigman)$$ and every observable
has the form $$A= a_0 +
\av\cdot\fat{$\sigma$},\equ(Asigma)$$ where $a_0$ is a real scalar and $\av$,
a real 3-vector.  A set of observables $A, B, C,\ldots$ is mutually
commuting if and only if the vectors $\av, \bv, \cv,\ldots$ are all
parallel. The eigenvalues of $A$, and hence the allowed values $v(A)$, are
restricted to the two numbers $$v(A) = a_0\pm a,\equ(eigen)$$ where $a$ is
the magnitude of the vector \av.  The simultaneous eigenvalues of a set of
mutually commuting observables are given by chosing one sign in \(eigen) for
those observables whose vectors point one way along their common direction,
and the opposite sign for those whose vectors point the other way.  Because
each observable $A$ takes on only two values the distribution of those
values in a given state is entirely determined by the mean of $A$, which is
given by $$\<\uparrow_n|A|\uparrow_n\> = a_0 +
\av\cdot\nv.\equ(mean)$$ 

A rule associating with each observable one of its eigenvalues will yield
simultaneous eigenvalues for mutually commuting observables if it always
specifies the opposite sign in \(eigen) for commuting observables associated
with oppositely directed vectors.  We require, in addition, for each state
$|\uparrow_n\>$, that the rule specify a distribution of those values
yielding the statistics demanded by \(mean).  Here is a rule that does
everything:\ft{It is a little simpler then the one Bell gives. One can
extend the rule to cover the case $(\mv+\nv)\cdot\av = 0$, but since this
has zero statistical weight, I do not bother.  Note that the values assigned
to non-commuting observables do not satisfy von Neumann's additivity
condition in individual members of the ensemble, although their average over
the ensemble does, which is all quantum mechanics requires.} Given a
particular individual system from an ensemble described by the state
$|\uparrow_n\>$, pick at random a second unit vector $\mv$ (which plays the
role of the hidden--variable) and assign to each observable $A$ the values
$$\eqalign{v_n(A) &= a_0 + a,\ {\rm if}\ (\mv+\nv)\cdot\av > 0,\cr
           v_n(A) &= a_0 - a,\ {\rm if}\ (\mv+\nv)\cdot\av <
0.}\equ(mvalue)$$ An elementary integration confirms that the mean over a
uniform distribution of directions of $\mv$ of the value \(mvalue) of any
observable in the state $|\uparrow_n\>$ is indeed given by the quantum
mechanical result
\(mean):  $$\int {d\Omega_m\over 4\pi}v_n(A) = a_0 +
\av\cdot\nv.\equ(meanhv)$$

\bigskip
\head{IV. THE BELL--KOCHEN--SPECKER THEOREM}\nobreak
\medskip\nobreak Having thus given an absurdly simple example of what had
solemnly been declared impossible for the past three decades, Bell proceeded
to show that the trick could no longer be accomplished in a state space of 3
or more\ft{His argument focuses on a space of exactly 3 dimensions, which
can, however, be a subspace of a higher dimensional space; the same remark
applies to the new arguments in 4 and 8 dimensions given in Sections V and
VI.} dimensions;\ft{Peculiar to 2 dimensions is the fact that all
observables that commute with any non-trivial observable $A$ necessarily
commute with each other.} \ie he gave a new \nhv\ proof that did not rely on
the silly condition.  I now give the full proof of this \bks\ theorem, but
here too, I include it only to emphasize the much greater simplicity of the
new versions that follow in Sections V and VI, to which readers with no
interest in the early history of the subject may jump without conceptual
loss. 

Just as it is convenient to use the algebra of spin--$\fr1/2$ to describe a
2-dimensional state space, it is also convenient to describe the
3-dimensional state space in terms of observables built out of angular
momentum components for a particle of spin-1.\ft{Bell actually works with
orthogonal projections, but the correspondence is entirely trivial:  $S_x^2
= 1 - P_x$, {\it etc.} I find it more congenial to follow Kochen and Specker
in using spin components, though the version of the argument I give is
Bell's, not theirs.} The observables we consider are the squares of the
components of the spin along various directions.  Such observables have
eigenvalues 1 or 0, since the unsquared spin components have eigenvalues 1,
0, or $-1$.  Furthermore the sums of the squared spin components along any
three orthogonal directions $u, v,$ and $w$ satisfy $$S_u^2 +S_v^2 +S_w^2 =
s(s+1) = 2,\equ(spin1)$$ since we are dealing with a particle of spin-1
($s=1$).  Finally the squared components of the spin along any three
orthogonal directions constitute a mutually commuting set.\ft{This is not a
general property of angular momentum components but it does hold for
spin--1, as is evident from the correspondence with orthogonal projections
noted in the preceding footnote.} 

Suppose we are given a set of directions containing many different
orthogonal triads, and the corresponding set of observables consisting of
the squared spin components along each of the directions.  Since the three
observables associated with any orthogonal triad commute, they can be
simultaneously measured and the values such a measurement reveals for each
of them, 0 or 1, must satisfy the same constraint \(spin1) as the
observables themselves.  Thus two of the values must be 1 and the third, 0.
We would have a \nhv\ theorem if we could find a quantum mechanical state in
which the statistics for the results of measuring any three observables
associated with orthogonal triads could not be realized by any distribution
of assignments of 1 or 0 to every direction in the set, consistent with the
constraint.

The \bks\ theorem does substantially more than that:  it produces a set of
directions for which there is no way whatever to assign 1's and 0's to the
directions consistent with the constraint \(spin1), thereby rendering the statistical
state-dependent part of the argument unnecessary.  This is accomplished by
solving the following problem in geometry: Find a set of 3-dimensional
vectors (\ie directions) with the property that it is impossible to color
each vector red (\ie assign the value 1 to the squared spin component along
that direction) or blue (\ie assign the value 0) in such a way that every
subset of 3 mutually orthogonal vectors contains just one blue and two red
vectors. 

The unpleasantly tedious part of the solution consists of showing that if
the angle between two vectors of different color is less than
$\tan^{-1}(0.5)$ = 26.565 degrees, then we can find additional vectors
which, with the original two, constitute a set that cannot be colored
according to the rules.  Since all that matters is the direction of each
vector, we can choose their magnitudes at our convenience.  We take the blue
vector to be a unit vector
\zv\ defining the $z$-axis and take the red vector \av\ to lie in the $y$-$z$
plane:  $\av = \zv + \alpha\yv,\ \ 0 < \alpha < 0.5$.

We now make several elementary observations:

\item{1.} Since \zv\ is blue, \xv\ and \yv\ must both be red.\ft{As I mention
each new vector, add it to the set.} 

\item{2.} Indeed, any vector in the $x$-$y$ plane must be red, since one
cannot have two orthogonal blue vectors.  In particular \cv\ = $\beta\xv +
\yv$ must be red, for arbitrary $\beta$.  Particularly interesting values of
$\beta$ will be specified shortly.

\item{3.} Similarly, since \av\ and \xv\ are red, any vector in their plane,
and, in particular, $\dv = \xv/\beta - \av/\alpha$ must be red.\ft{If you
happen to be interested in counting how many vectors are in the uncolorable set we
end up with, then whenever we add a red vector {\bf v} in the plane of two
orthogonal red vectors you should also add to the set, if they are not
already present, a second red vector in that plane perpendicular to {\bf v},
as well as a blue vector perpendicular to the plane.}

\item{4.} Because \av\ = \zv\ + $\alpha$\yv, \dv\ is orthogonal to \cv =
$\beta$\xv+\yv .  Since both \cv\ and \dv\ are red, the normal to their
plane must be blue, and therefore any vector in their plane, in particular
\ev\ = \cv\ + \dv\ must be red. 

\item{5.} But adding the explicit forms of \cv\ and \dv\ we see that $\ev
= (\beta + \beta^{-1})\xv - \zv/\alpha.$

\item{6.} Since $\alpha$ is less than 0.5, $1/\alpha$ is greater than 2.
Since $|\beta + \beta^{-1}|$ ranges between 2 and $\infty$ as $\beta$ ranges
through all real numbers, we can find a value of $\beta$ such that \ev\ is
along the direction of $\fv\ =
\xv - \zv$.  Changing the sign of $\beta$ gives another \ev\ along the
direction of $\gv = -\xv - \zv$.

\item{7.} Since \ev\ is red whatever the value of $\beta$, \fv\ and \gv\ must
be red.

\item{8.} But \fv\ and \gv\ are orthogonal.  The normal to their plane is
therefore blue and any vector in their plane is necessarily red.

\item{9.} But $\zv = -\fr1/2\fv - \fr1/2\gv$ is in the plane of \fv\ and
\gv, and $\zv$ is blue.

\item{10.} Contradiction!  Therefore the set cannot be colored according to
the rules if \av\ and \zv\ have different colors. 

The rest is genuinely trivial.  We find an uncolorable set of directions by
noting that since $22.5$ degrees $< \tan^{-1}(0.5)$, the $z$-axis must have
the same color as a direction $22.5$ degrees away from it in the $y$-$z$
plane, or we could produce an uncolorable set as described above.  But that
direction must then have the same color as the direction in the $y$-$z$
plane another $22.5$ degrees away from the $z$-axis.  Two more such steps
gets us down to the $y$-axis, which must thus have the same color as the
$z$-axis.  Repeating this procedure in the $y$-$x$ plane we conclude that
the $x$-axis must share that same color.  But three mutually orthogonal axes
cannot all have the same color: two must be red and one blue.  Therefore the
5 directions in the $y$-$z$ plane plus the 4 additional directions in the
$x$-$y$ plane plus the additional directions needed to carry out steps 1-10
above for each pair separated by $22.5$ degrees constitute an uncolorable
set. 

Bell did not conclude his proof with these elementary remarks about steps of
22.5 degrees.  Instead, after proving that differently colored directions
must be more than a minimum angle apart, he simply noted that it was
therefore impossible to associate a color with every direction, since any
coloring of the sphere with just two colors obviously must have different
colors arbitrarily close together.  As a result, many philosophers
characterize his proof as a ``continuum proof'' and prefer the argument
Kochen and Specker independently gave a year later, which gives a slightly
different (weaker) version of the minimum angle theorem but explicitly
displays a finite set of directions---117 of them---which cannot be
colored according to the rules.  Clearly the Bell argument as stated above
also uses only a finite set of directions.  But there is no point making a
fuss about this because both arguments have now been superseded by an
argument that is also state--independent, whose algebraic part is even more
elementary (appealing to no possibly unfamiliar result about the commutation
of squares of orthogonal spin components), which requires no subsequent
geometric analysis at all, and which uses far fewer observables.

The only price one pays for the simplicity is that the argument now requires
a state space of at least 4 dimensions.\ft{In hindsight this might have been
guessed: if a \nhv\ theorem is impossible in 2 dimensions, and rather
complicated in 3, extrapolation suggests that it might be easy in 4.} So
unless one has a special interest in proving \nhv\ theorems in 3 dimensions,
one can safely declare the old Bell or Kochen-Specker versions of the theorem
obsolete, sparing future generations of philosophers of science a
painful rite of passage and making the result readily available even to
physicists in ten minutes of an introductory quantum mechanics course.

Before consigning the old argument to the history books, I digress to remark
that the 117 directions seem to have held a great power over the philosophic
imagination. 
The cover illustration of a well-known
treatise on the philosophy of quantum mechanics,\ftn*{Michael Redhead,
1987.   The cover is reproduced as Figure 1 in Mermin, 1993.} for example, is emblazoned with an intricate diagram used by Kochen
and Specker, 1987, to represent their set of 117 uncolorable directions.
Although the diagram is unfamiliar to all but a handful of quantum
physicists, a distinguished philosopher of science regarded it as an
appropriate icon for the entire subject.  

Since 1967 other sets of uncolorable directions have been discovered with
fewer vectors.  The current world record holders are J. Conway and S.
Kochen\ft{S. Kochen, private communication.} with 31, but Asher Peres, 1991,
has found a prettier set of 33 with cubic symmetry which can be exploited to
give a proof of the no-coloring theorem that is more compact than
Bell's.  Roger Penrose has pointed out that Peres's set of 33 directions can be
described as follows: take a cube and superimpose it with its 90 degree
rotations about two perpendicular lines connecting its center to the
midpoints of an edge.  Peres's directions point to the vertices and to the
centers of the faces and edges of the resulting set of 3 interpenetrating
cubes.  This very figure occurs as a large ornament atop one of the two
towers in M.~Escher's famous drawing of the impossible waterfall.\ftn{*}{Escher, 1970.  The waterfall is reproduced as Figure 2 of Mermin, 1993.} 

\bigskip
\head{V. A SIMPLER BELL-KS THEOREM IN 4 DIMENSIONS}
\medskip

I now turn to the version of the Bell-KS theorem that works in a
4-dimensional state space.\ft{This argument was inspired by an earlier
version by A.~Peres, 1990, that uses an even smaller number of observables,
but only applies to an ensemble prepared in a particular state.} Our task is
exactly the same as Bell, Kochen, and Specker faced in the 3-dimensional
case: we must exhibit a set of observables $A$, $B$, $C\ldots$ for which we
can prove that it is impossible to associate with each observable one of its
eigenvalues, $v(A), v(B), v(C),\ldots$ in such a way that all functional
relationships between mutually commuting subsets of the observables are also
satisfied by the associated values.  The only difference is that now we can
do the trick with many fewer observables, and an entirely trivial proof.

In 4 dimensions it is convenient to represent observables in terms of the
Pauli matrices for two independent spin-$\fr1/2$ particles:\ft{These are
simply to be viewed as a convenient set of operators in terms of which to
expand more general 4-dimensional operators; we need not be talking about
two spin-$\fr1/2$ particles at all.} $\sa1\mu$ and $\sa2\nu$.  The relevant
properties of these observables are the familiar ones: the squares of each
are unity so the eigenvalues of each are $\pm1$; any component of $\sa1\mu$
commutes with any other component of $\sa2\nu$; when $\mu$ and $\nu$ specify
orthogonal directions $\sa i\mu$ anticommutes with $\sa i\nu$ for $i=1,2$;
and $\sa ix\sa iy = i\sa iz$ for $i=1,2$.  Consider, then, the nine
observables shown below, which it is convenient to arrange in groups 
of three on six intersecting lines that form a square.  

\def\sa#1#2{\sigma^#1_#2}

\def\sc#1#2#3{\sigma^1_#1\sigma^2_#2\sigma^3_#3}

\def\ha{\hskip0.15truein}
\def\ch#1{\centerline{\hbox{#1}}}
\def\vs#1{\vskip#1pt}
\def\hs#1{\hskip#1truein}
\bigskip
{\baselineskip=0pt
\ch{$\sa1x        \hs{0.5}       \sa2x        \hs{0.4}    \sa1x\sa2x $} 
\vs{20}
\ch{$\sa2y        \hs{0.5}        \sa1y       \hs{0.4}    \sa1y\sa2y $}
\vs{20} 
\ch{$ \sa1x\sa2y  \hs{0.3}   \sa2x\sa1y       \hs{0.3}    \sa1z\sa2z $}     
}
\bigskip

To prove that it is
impossible to assign values to all nine observables we merely note that:

 \item{(a)} The observables in each of the three rows and each of the three
columns are mutually commuting.  This is immediately evident for the top two
rows and first two columns from the left; it is true for the bottom row
and right-hand column because in every case there is a pair of
anti-commutations.

\item{(b)} The product of the three observables in the column on the right is
$-1$.  The product of the three observables in the other two columns and all
three rows is $+1$.

\item{(c)} Since the values assigned to mutually commuting observables must
obey any identities satisfied by the observables themselves, the identities
(b) require the product of the values assigned to the three observables in
each row to be 1, and the product of the values assigned to the three
observables in each column to be 1 for the first two columns and $-1$ for
the column on the right.  

\noindent But (c) is impossible to satisfy, since the row identities require
the product of all nine values to be 1, while the column identities require
it to be $-1$.

I maintain that this is as simple a version of the Bell-KS theorem as one is
ever likely to find,\ft{Peres, 1991, recasts the argument as a no-coloring
theorem for a set of 24 directions in 4--dimensions, thereby making it
complicated again.  The advantage of the 4-dimensional argument over the
traditional one in 3 dimensions is just that no such analysis is
necessary.} and that it belongs in elementary texts on quantum mechanics as
a direct demonstration, straight from the formalism, without any appeal to
decrees by the Founders, that one cannot realize the naive ensemble
interpretation of the theory on which the attempt to assign values is
based.  It is nevertheless susceptible to the same criticism that Bell
himself immediately brought to bear against his own version of the theorem.
Before turning to that criticism, however, I describe a comparably simple
version of the \bks\ theorem that works in an 8-dimensional state
space\ft{That the 3-spin form of the Greenberger-Horne-Zeilinger version
of Bell's Theorem could be reinterpreted as a version of the \bks\ theorem
was brought to my attention by A. Stairs.} that we shall find is capable of
evading Bell's criticism in a way that the 4-dimensional version is not.
The 8-dimensional argument provides a direct link between the \bks\
theorems and their illustrious companion, Bell's Theorem, when Bell's
theorem is presented in the spectacular form recently discovered by
Greenberger, Horne, and Zeilinger, 1989. 

\vfil\eject

\bigskip
\head{VI. A SIMPLE AND MORE VERSATILE Bell-KS THEOREM} 
\head{IN 8 DIMENSIONS.}
\medskip

We construct our 8-dimensional observables out of three independent
spins--$\fr1/2$, and consider the set of ten observables shown below, 
which it is now convenient to arrange in groups of four on 5 intersecting
lines that form a five-pointed star: 

\def\sa#1#2{\sigma^#1_#2}

\def\sc#1#2#3{\sigma^1_#1\sigma^2_#2\sigma^3_#3}
\def\ha{\hskip0.15truein}
\def\ch#1{\centerline{\hbox{#1}}}
\def\vs#1{\vskip#1pt}
\bigskip
{\baselineskip=0pt
\ch{$ \sa1y $} 
\vs{20}
\ch{$ \sc xxx \ha \sc yyx \ha \sc yxy \ha \sc xyy $}
\vs{15}                                               
\ch{$ \sa3x \hskip1truein \sa3y $}                    
\vs{3}                                                
\ch{$ \sa1x $} 
\vs{15}                                               
\ch{$ \sa2y \hskip1.8truein \sa2x $}                  
}
\bigskip 

To prove that it is impossible to
assign values to all 10 observables note that:

\item{(a)} The four observables on each of the five lines of the star are
mutually commuting.  This is immediately evident for all but the horizontal
line, where it follows from the fact that interchanging the observables in
each of the six possible pairs always requires a pair of anticommutations. 

\item{(b)} The product of the four observables on every line of the star but
the horizontal line is 1.  The product of the four observables on the
horizontal line is $-1$.

\item{(c)} Since the values assigned to mutually commuting observables must
obey any identities satisfied by the observables themselves, the identities
(b) require the product of the values assigned to the four observables on the
horizontal line of the star to be $-1$, and the product of the values
assigned to the four observables on each of the other lines to be $+1$.

\noindent Condition (c) requires the product over all five lines of the
products of the values on each line to be $-1$.  But this is impossible, for
each observable is at the intersection of two lines.  Its value appears
twice in the product over all five lines, and that product must therefore be
+1.  

This hardly more elaborate 8-dimensional version of the theorem has an
additional virtue that the 4-dimensional version lacks.  To see this and to
see the connection with Bell's Theorem we turn, finally, to Bell's
objection to his own argument. 

\bigskip
\head{VII. IS THE Bell-KS THEOREM SILLY?}
\medskip

In all these cases, as Bell pointed out immediately after proving the \bks\
theorem, we have ``tacitly assumed that the measurement of an observable
must yield the same value independently of what other measurements must be
made simultaneously.'' In Bell's 3-dimensional example and in both the 4-
and 8-dimensional examples we required each observable to have a value in
an individual system that would give the result of its measurement, {\it
regardless of which of two sets of mutually commuting observables we chose
to measure it with.\/} But since the additional observables in one of those
sets do not all commute with the additional observables in the other, the
two cases are incompatible.  ``These different possibilities require
different experimental arrangements; there is no {\it a priori\/} reason to
believe that the results $\ldots$ should be the same.  The result of an
observation may reasonably depend not only on the state of the system
(including hidden-variables) but also on the complete disposition of the
apparatus.''\ftn*{Bell, 1966.} 

This tacit assumption that a hidden-variables theory has to assign to an
observable $A$ the same value whether $A$ is measured as part of the 
mutually commuting set $A, B, C,\ldots$ or a second mutually commuting set
$A, L, M,\ldots\,$ even when some of the $L, M,\ldots$ fail to commute with
some of the $B, C,\,\ldots$, is called ``non-contextuality'' by the
philosophers.  Is non-contextuality, as Bell seemed to suggest, as silly a
condition as von Neumann's---a foolish disregard of ``the impossibility of
any sharp distinction between the behavior of atomic objects and the
interaction with the measuring instruments which serve to define the
conditions under which the phenomena appear,'' as Bohr\ft{N. Bohr in Schilpp,
1949 and cited in Bell, 1966.  Bell's invocation of Bohr, to whom any
hidden-variables theory would have been anathema, in order to dismiss the
implications of his own \nhv\ theorem, thereby maintaining the viability of
the hidden-variables program, was aptly characterized by Abner Shimony as
``a judo-like maneuver.''} put it?

I would not characterize the assumption of non-contextuality as a silly
constraint on a hidden-variables theory.  It is surely an important fact
that the impossibility of embedding quantum mechanics in a non-contextual
hidden-variables theory rests not only on Bohr's doctrine of the
inseparability of the objects and the measuring instruments, but also on a
straightforward contradiction, independent of one's philosophic point of
view, between some quantitative consequences of non-contextuality and the
quantitative predictions of quantum mechanics.

Furthermore, there are features of quantum mechanics that seem strongly to
hint at an underlying non-contextual \hvt\ as the only available
explanation.\ft{An ``only available explanation'' is one to which the only
alternative is the claim that no explanation is required.} Most strikingly,
although it is indisputable that measuring $A$ with mutually commuting $B,
C,\ldots$ requires a different experimental arrangement from measuring it
with mutually commuting $L, M,\ldots$ whenever some of $L, M,\ldots$ fail to
commute with some of $B, C,\ldots\,$, it is nevertheless an elementary theorem
of quantum mechanics that the joint distribution $p(a,b,c,\ldots)$ for the
first experiment yields precisely the same marginal distribution $p(a)$ as
does the joint distribution $p(a,l,m\ldots)$ for the second, in spite of the
different experimental arrangements.  If we do the experiment to measure $A$
with $B, C,\ldots$ on an ensemble of systems prepared in the state $\Psi$
and ignore the results of the other observables, we get {\it exactly the
same\/} statistics for $A$ as we would have obtained had we instead done the
quite different experiment to measure $A$ with $L, M, \ldots$ on that same
ensemble.  The obvious way to account for this, particularly when
entertaining the possibility of a \hvt, is to propose that both experiments
reveal a set of values for $A$ in the individual systems that is the same,
regardless of which experiment we choose to extract them from.  Putting it
the other way around, a {\it contextual\/} hidden-variables account of this
fact would be as mysteriously silent as the quantum theory on the question
of why nature should conspire to arrange for the marginal distributions to
be the same for the two different experimental arrangements.

Of course if the method to measure $A$ with mutually commuting $B, C,\ldots$
consists of successive filtrations---first measure $A$, then $B$, then $C$,
{\it etc.}--- and successive filtrations are also used to measure $A$ with
mutually commuting $L, M,\ldots$, then if $A$ is the first observable tested
in either case, the resulting statistics for $A$ alone will necessarily be
the same in both cases, since we need not even decide which case to proceed
with until after we have acquired the result of the $A$-measurement.  But
this merely shifts the puzzle raised by the non-contextuality of quantum
mechanical probabilities to a new form: why should the statistical results
of a sequential measurement of a set of mutually commuting observables be
independent of the way we order them?  Even more puzzling, why are those
statistics unaffected if we change to quite a different way of determining
them?  We could, for example, measure three mutually commuting observables
$A$, $B$, and $C$, each with eigenvalues 1 or 0 (like the squared spin
components in the original \bks\ argument) by measuring the single
observable $4A+2B+C$, the three digit binary form of the result giving
precisely the values of $A$, $B$, and $C$.  If one is attempting a
hidden-variables model at all, it seems not unreasonable to expect the model
to provide the obvious explanation for this striking insensitivity of the
distribution to changes in the experimental arrangement---namely that the
hidden variables are non-contextual.

There is, however, one class of \nhv\ theorems in which non-contextual\-ity
can be replaced by an even more compelling assumption, which brings us,
finally, to Bell's Theorem (Bell, 1964).  

\bigskip
\head{VIII. LOCALITY REPLACES non-contextuality:}
\head{ BELL'S THEOREM}
\medskip

Suppose that the experiment that measures commuting observables $A$, $B$,
$C$, $\ldots$ uses independent pieces of equipment far apart from one
another, which separately register the values of $A$, $B$, $C$, $\ldots\,$.
And suppose that the experiment to measure $A$ with the commuting observable
$L$, $M$, $\ldots\,$, not all of which commute with all of $B$, $C$,
$\ldots\,$, requires changes in the complete apparatus that amount only to
replacing the parts that register the values of $B$, $C$, $\ldots$ with
different pieces of equipment that register the values of $L$, $M$,
$\ldots\,$.  And suppose that all these changes of equipment are made far
away from the unchanged piece of apparatus that registers the value of $A$.
In the absence of action at a distance such changes in the complete
disposition of the apparatus could hardly be expected to have an effect on
the outcome of the $A$--measurement on an individual system.  In this case
the problematic assumption of non-contextuality can be replaced by a
straightforward assumption of locality.

Can we prove a Bell-KS theorem in which we only assume non-contextuality
when it can be justified by locality?  I know of no way to accomplish this
trick that works for arbitrary states, but if one is willing to settle for a
proof that works only for suitably prepared states, then it can easily be
done.  This was first accomplished in Bell's Theorem, which in its original
form applies to a pair of far-apart spin-$\fr1/2$ particles {\it in the
singlet state.\/} An analogous theorem can be established by a very minor
modification of the 8-dimensional version of the \bks\ theorem.\ft{The
modification converts it into the model of Greenberger, Horne, and
Zeilinger, in the version I gave in Mermin, 1990b,c.} This new version of
Bell's Theorem makes it clear that the use of a particular state is required
to provide the information that is lost when one permits the assignment of
non-contextual values only when non-contextuality is a consequence of
locality.

To convert the 8-dimensional version of the \bks\ theorem into a form of
Bell's Theorem, we interpret the three vector operators $\fat{$\sigma$}^i$,
until now merely a convenient set from which to construct more general
observables, as literally describing the spins of three different
spin-$\fr1/2$ particles, localized far away from one another.  An
examination of the ten observables appearing in Fig. 4 reveals that all but
the four appearing on the horizontal line of the star describe spin
components of a single isolated particle.  Setting aside the four non-local
observables, each of which is built out of the product of spin components of
all three particles, we are left with six observables belonging to four
sets, each containing three local observables, lying on the four
non-horizontal lines of the star.  Each observable associated with a single
particle appears in two of these sets, which differ in the selection of the
pair of observables associated with the two far away particles.  For any of
these six local observables, the assumption that the value assigned it
should not depend on which pair of far-away components are measured with it
is justified not by a possibly dubious assumption of non-contextuality, but
by the condition of locality.  

By dropping the non-contextual assignment of values to the four non-local
observables, however, we break the chain of relations that led to a
contradiction in the \bks\ argument.  We can rescue the argument by noting
that because all four non-local observables {\it commute\/} with each other,
they have simultaneous eigenstates.  In an ensemble of individual systems
prepared in such an eigenstate, the non-local observables all have definite
values for valid and conventional quantum mechanical reasons.  These values
play the same role in the new argument as the non-contextual values assigned
them played in the earlier version, being related to the values of the
appropriate sets of three local observables in exactly the same way.\ft{For
example if $\sa 1x, \sa 2x$, and $\sa 3x$ are measured in an eigenstate of
$\sc xxx$ with given eigenvalue, orthodox quantum mechanics requires the
product of the three results to be equal to that eigenvalue.} The only
difference is that because we now consider systems in an {\it eigenstate\/}
of all four non-local observables, those four simultaneous values cannot
fluctuate among the eight possible sets they might in general possess, but
are fixed to a set of values.  This further constraint does not alter the
conclusion that there is no consistent way to assign values to all ten
observables and thus, in particular, no consistent assignment of values to
the six local observables.

The 8-dimensional model of three spins--$\fr1/2$ therefore provides a
conceptual link between the two theorems of John Bell that was not evident
in their original forms.  The difference between the two 8-dimensional
arguments is that the
\bks\ version rules out the assignment of non-contextual values to arbitrary
observables, while the \bt\ version rules it out even when
non-contextuality is restricted to cases where it can be justified on the
basis of locality.  While both theorems demonstrate that the assignment is
impossible, the demonstration based on locality is the more powerful result
since it applies even under a restricted use of non-contextuality.  

Because the \bks\ version applies to
\nhv\ theories that are allowed to assign non-contextual values to a more
general class of observables than in the Bell's Theorem version, the \bks\
version does not need the properties of a particular state.  In Bell's
original versions of these theorems, where the arguments could not be set
side--by--side, this appeared to be a compensating strength of the
\bks\ argument.  In the new version, however, it is seen to be merely a
technical consequence of the fact that by making a broader assignment of
non-contextual hidden variables, the \bks\ argument can dispense with one of
the stratagems the more powerful argument of Bell's Theorem requires to
produce its counterexample.

It is instructive to see why we cannot convert the 4-dimensional version of
the
\bks\ theorem into an argument based on locality.  In that argument (see
Fig.~3) there are four local and five non-local observables that we now
interpret as describing two far-apart spin-$\fr1/2$ particles.  Each local
observable can be measured with either of two other local observables that
fail to commute with each other, associated with the other far-away
particle.  If we wish to make the asumption of non-contextuality only when
it is required by the weaker assumption of locality, then we cannot assign
non-contextual values to the five non-local observables, and need some
other way to complete the chain leading to a contradiction.  But in contrast
to the 8-dimensional argument, the non-local observables do not all
commute.  It is thus no longer possible to assign values to all five by
considering an ensemble of systems prepared in a simultaneous eigenstate.
The theorem cannot be converted into a version of \bt.

Note that locality can be used not only to justify the condition of
non-contextuality, but also to motivate further the attempt to assign values
to the local observables in the first place.  For in an ensemble of systems
described by a simultaneous eigenstate of the non-local observables, the
results of measuring any one of the local observables on an individual
system can be determined prior to the measurement, by first measuring far
away an appropriate set of two other local observables. Because the results
of the measurements of the three local observables must be consistent with
the eigenvalue of the observable that is their product, any two such results
determines the third.  As noted long ago by Einstein, Podolsky, and Rosen
(Einstein, 1935), in the absence of spooky actions at a distance it is hard
to understand how this can happen unless the two earlier measurements are
simply revealing properties of the subsequently measured particle that
already exist prior to their measurement.

\bigskip
\head{IX. A LITTLE ABOUT BOHM THEORY}
\medskip

Bell's favorite example of a hidden-variables theory, Bohm theory,\ftn*{Bohm,
1952.} is not only explicitly contextual but explicitly and spectacularly
non-local,\ft{This is noted in Bell, 1966, where Bell raises the question
of whether ``{\it any\/} hidden-variables account of quantum mechanics {\it
must\/} have this extraordinary character.  (Remember, this was written
before Bell, 1964!)  Bell, 1982, reprinted as Chapter 17 of Bell, 1987,
gives a more detailed discussion of Bohm theory from this perspective.
Chapters 14 and 15 of Bell, 1987 give an exceptionally clear and concise
exposition of Bohm theory.} as it must be in view of the Bell-KS theorem
and Bell's Theorem.  In Bohm theory, which defies all the impossibility
proofs, the hidden variables are simply the real configuration space
coordinates of real particles, guided in their motion by the wave function,
which is viewed as a real field in configuration space.  The wave function
guides the particles like this:\ft{I describe only spinless particles, but
spin can also be handled.} each particle obeys a first order equation of
motion specifying that its velocity is proportional to the gradient with
respect to its position coordinates of the phase of the $N$-particle wave
function, {\it evaluated at the instantaneous positions of all the other
particles.} It is the italicized phrase which is responsible for the
``hideous'' non-locality whenever the wave function is correlated.\ft{If
the wave function factors then the phase is a sum of phases associated with
the individual particles and the non-locality goes away.} One easily proves
that if the wave function obeys Schroedinger's equation, then a distribution
of initial coordinates of the particles given by $|\Psi_0|^2$, will evolve
under these dynamics into $|\Psi_t|^2$ at time $t$.\ft{This is the way
Bell presents Bohm theory.  Bohm prefers to take another time derivative of
the equation of motion for the particles to make it look more like $F=ma$,
which he gets, with corrections to the classical force arising from what he
calls the ``quantum potential.''}

If two particles are in a correlated state then because the field guiding
the second particle depends on the trajectory of the first, if a field is
suddenly turned on in a region where the first particle happens to be, the
subsequent motion of the second particle can be drastically altered in a
manner that does not diminish with the distance between the two particles.
Since measurements on each of a collection of non-interacting particles can
be described by the action of just such fields, this gives contextuality
with a vengeance.

\bigskip
\head{X. THE LAST WORD} 
\medskip

John Bell did not believe that either of his \nhv\ theorems excluded the
possibility of a deeper level of description than quantum mechanics, any
more than von Neumann's theorem does.  He viewed them all as identifying
conditions that such a description would have to meet.  Von Neumann's
theorem established only that a \hvt\ must assign values to non-commuting
observables that do not obey in individual systems the additivity condition
they satisfy in the mean---a result already evident from the trivial example
of $\sx+\sy$.  The
\bks\ theorems establish that in a \hvt\ the values assigned even to a set
of mutually commuting observables must depend on the manner in which they
are measured---a fact that Bohr could have told us long ago (although he
would have disapproved of the whole undertaking).  And Bell's Theorem
establishes that the value assigned to an observable must depend on the
complete experimental arrangement under which it is measured even when two
arrangements differ only far from the region in which the value is
ascertained---a fact that Bohm theory exemplifies, and which is now
understood to be an unavoidable feature of any hidden variables-theory.

To those for whom non-locality is anathema, Bell's Theorem finally spells
the death of the hidden-variables program.\ft{Many people contend that
\bt\ demonstrates non-locality independent of a hidden-variables program,
but there is not general agreement about this.} But not for Bell.  None of
the \nhv\ theorems persuaded him that hidden variables were impossible.
What Bell's Theorem did suggest to Bell was the need to reexamine our
understanding of Lorentz invariance, as he argues in his delightful essay on
how to teach special relativity\ftn*{Bell, 1987, p.~12.} and in Dennis Weaire's
transcription of Bell's lecture on the Fitzgerald contraction.\ftn{**}{Bell,
1992.} ``What is proved by impossibility proofs,'' Bell declared, ``is lack
of imagination.''\ft{Bell, 1982.  Although I gladly give John Bell the last
word, I will take the last footnote to insist that he is unreasonably
dismissive of the importance of his own impossibility proofs.  One could
make a complementary criticism of much of contemporary theoretical physics:
What is proved by possibility proofs is an excess of imagination.  Either
criticism undervalues the importance of defining limits to what speculative
theories can or cannot be expected to accomplish.}

\bigskip
\head{ACKNOWLEDGMENT} 
\medskip

Supported by National Science Foundation under Grant No.~PHY 9022796.  This
is a revised and expanded version of the text of the Bell Memorial Lecture
given at the XIXth International Colloquium on Group Theoretic Methods in
Physics, Salamanca, July, 1992.  (The earlier version is to appear in the
proceedings of the Salamanca conference.)  My treatment of these issues
evolved through half a dozen general physics colloquia, given during
academic 1991-92, and has benefitted from the thoughtful responses of
skeptical members of those audiences.  Many people contributed to my
formulation and discussion of the new versions of both Bell theorems, with
clever ideas, wise criticisms, or instructive failures to grasp points I
foolishly thought I had made with transcendent clarity.  I am especially
indebted to Harvey Brown, Robert Clifton, Anthony Garrett, Kurt Gottfried,
Daniel Greenberger, Jon Jarrett, Asher Peres, Abner Shimony, and Alan Stairs.

This essay is dedicated to the memory of my brother Joel Mermin (1939--1992)
who loved to take long walks and simplify theorems.  

\medskip
\head{REFERENCES}
\medskip

{
\baselineskip=13.7pt
\parindent=0pt
\parskip=5pt

Adler, Mortimer J., 1992, ``Natural Theology, Chance, and God,'' {\it The
Great Ideas Today\/}, Encyclopoedia Britannica, Chicago, 288--301.

Bell, J. S., 1964, ``On the Einstein-Podolsky-Rosen Paradox,'' Physics
{\bf 1}, 195-200.

Bell, J. S., 1966, ``On the problem of hidden variables in quantum
mechanics,'' Revs.  Mod. Phys. {\bf 38}, 447-52.

Bell, J. S., 1982, ``On the impossible pilot wave,'' Foundations of Physics
{\bf 12}, 989-999.

Bell, J. S., 1987, {\it Speakable and Unspeakable in Quantum Mechanics\/},
Cambridge University Press, Cambridge.

Bell, J. S., 1992, ``George Francis Fitzgerald'', Physics World {\bf 5}, No.
9, 31-35.  (Lecture at Trinity College, Dublin, 1989, as transcribed by
Dennis Weaire.)

Bohm, D., 1952,``A suggested interpretation of the
quantum theory in terms of `hidden variables',''  Phys. Rev. {\bf 85},  66-179 (Part I), 
180-193 (Part II).

Bub J., 2010. ``Von Neumann's 'No hidden variables' proof: A re-appraisal.'' Foundations of Physics, {\bf 40},  1333-1340.

Bub J., 2011. ``Is Von NeumannÕs `no hidden
variables' proof silly?", Chapter 10 of {\it Deep beauty --- 
understanding the quantum world through mathematical innovation'\/}, H. Halvorson, ed., Princeton University Press, 
393-407.

Dieks, D., 2017, ``Von Neumann's impossibility proof: Mathematics
in the service of rhetorics, Studies in the History and Philosophy of Modern Physics,    136-148.  

Einstein, A., Podolsky, B., and Rosen, N., 1935, ``Can quantum--mechanical
description of physical reality be considered complete?'' Phys. Rev. {\bf
47}, 777-780.

Escher, M., 1970, {\it The Graphic Work of M.~C.~Escher\/}, Hawthorn Books,
New York, Plate 76, ornament atop the left tower.

Gleason, A.M., 1957, ``Measures on the closed subspaces of a Hilbert
space,'' J. Math.  Mech. {\bf 6}, 885-893.

Greenberger, D.M., M.Horne, and A.  Zeilinger, 1989, ``Going beyond Bell's
theorem,'' in {\it Bell's Theorem, Quantum theory, and Conceptions of the
Universe,} ed. M.  Kafatos (Kluwer, Dordrecht), pp.  73-76. 

Greenberger, D.M., M. A. Horne, A.  Shimony, and A.  Zeilinger, 1990,
``Bell's theorem without inequalities,'' Am. J. Phys. {\bf 58}, 1131-1143.

Hermann, G., 1935, ``Die naturphilosophischen Grundlagen der Quantenmechanik
(An\-zug),'' Abhandlungen der Freis'schen Schule {\bf 6}, 75-152.

Jammer, M., 1974 {\it The Philosophy of Quantum Mechanics\/}, Wiley,
New York, p. 273.

Kochen, S. and E. P. Specker, 1967, ``The problem of hidden variables in
quantum mechanics,'' J. Math. Mech. {\bf 17}, 59-87.

Mermin, N. D., 1990a, {\it Boojums All the Way Through}, Cambridge
University Press, New York, Chapters 10--12.

Mermin, N. D., 1990b, ``Simple unified form for the major
no--hidden-variables theorems,'' Phys.  Rev.  Lett. {\bf 65}, 3373-7.

Mermin, N. D., 1990c, ``What's wrong with these elements of reality?'' Phys.
Today {\bf 43}(6), 9.

Mermin, N. D., 1990d, ``Quantum mysteries revisited,'' Am. J. Phys. {\bf
58}, 731-734. 

Mermin, N. D., 1993, ``Hidden variables and the two theorems of John Bell,'' Revs. Mod. Phys. {\bf 65}, 803-815.  

Peres, A, 1990, ``Incompatible results of quantum measurements,'' Phys.
Lett. A{\bf 151}, 107-108.

Peres, A., 1991, ``Two Simple Proofs of the Kochen--Specker Theorem,''
J. Phys. A {\bf 24} L175-178.  


Redhead, M., 1987, {\it Incompleteness, Nonlocality, and Realism,}
Clarendon Press, Oxford.  

Schilpp, P. A. ed., 1949, {\it A.  Einstein, Philosopher
Scientist\/}, Library of Living Philosophers, Evanston,
Ill.

von Neumann, J., 1932 {\it Mathematische Grundlagen der Quanten-mechanik\/},
Springer-Ver\-lag, Berlin.  English translation:\ \  {\it Mathematical Foundations
of Quantum Mechanics\/}, Princeton University Press, Princeton, N.J., 1955.
}


\bye
\vfil
\eject
\head{FIGURE CAPTIONS}

{\sl Figure 1.} The cover of Redhead, 1987, which reproduces the figure used
in the proof of the \bks\ theorem in Kochen and Specker, 1967.  The figure
contains 120 vertices representing 120 directions in 3--space, but the pairs
of directions $a,p_0$,\ $b,q_0$,\ and $c,r_0$ are the same, leaving the
notorious set of 117 distinct directions.  Reproduced by permission of
Oxford University Press.

\bigskip

{\sl Figure 2.} The tower on the left of M.~C.~Escher's engraving
``Waterfall.'' 
\copyright\ M.~C.~ Escher / Cordon Art -- Baarn -- Holland.  The ornament
atop the tower consists of three superimposed cubes.  One of the cubes has
all its edges horizontal or vertical.  The other two are given by rotating
this one through 90 degrees about each of the two perpendicular horizontal
lines that connect the midpoints of opposite vertical edges.  The 33
uncolorable directions used in the proof of the \bks\ theorem in Peres,
1991, lie along the lines connecting the common center of the cubes to their
vertices and the centers of their edges and faces.

\bigskip

{\sl Figure 3.} Nine observables leading to a very economical proof of the
\bks\ theorem in a state space of 4 or more dimensions.  The observables are
arranged in six groups of three, lying along three horizontal and three
vertical lines.  Each observable is associated with two such groups.  The
observables within each of the six groups are mutually commuting, and the
product of the three observables in each of the six groups is +1 except for
the vertical group on the right, where the product is $-1$.  

\bigskip

{\sl Figure 4.} Ten observables leading to a very economical proof of the
\bks\ theorem in a state space of 8 or more dimensions.  The observables are
arranged in five groups of four, lying along the five legs of a 5--pointed
star.  Each observable is associated with two such groups.  The observables
within each of the five groups are mutually commuting, and the product of
the four observables in each of the five groups is +1 except for the group
of four along the horizontal line of the star, where the product is $-1$.

\bye


\ft{Although boringly commonplace, this last assertion hides some very
subtle issues, which are exposed when one asks why, even if $A$ and $B$ do
not commute, measuring $A$ and then immediately measuring $B$ does not
qualify as a ``simultaneous measurement''?  One conventional answer is that
if they fail to commute, then the original measurement of $A$ disturbs the
system with uncontrollable consequences for the result of an immediate
subsequent measurement of $B$.  This seems to suggest that if $A$ and $B$
{\it do} commute, then the measurement of $A$ does {\it not} disrupt the
subsequent measurement of $B$, although it is not entirely clear what might
be meant by this last assertion if the initial state of the system is not a
simultaneous eigenstate of $A$ and $B$.  What one can say quite generally is
that if $A$ and $B$ do commute then the joint probability for the results of
simultaneously measuring $A$ and $B$ on a system in the state $|\psi\>$ does
not depend on how one carries out that simultaneous measurement; for example
the probability of a given pair of results is exactly the same whether one
measures $A$ and then immediately $B$, or the other way around.  This is not
the case for arbitrary states if $A$ and $B$ fail to commute.  We shall
return to this point below.} 

\def\ha{\hskip0.15truein}
\def\ch#1{\centerline{\hbox{#1}}}
\def\vs#1{\vskip#1pt}
\def\hs#1{\hskip#1truein}
\bigskip
{\baselineskip=0pt
\ch{$\sa1x        \hs{0.5}       \sa2x        \hs{0.4}    \sa1x\sa2x $} 
\vs{20}
\ch{$\sa2y        \hs{0.5}        \sa1y       \hs{0.4}    \sa1y\sa2y $}
\vs{20} 
\ch{$ \sa1x\sa2y  \hs{0.3}   \sa2x\sa1y       \hs{0.3}    \sa1z\sa2z $}     
}
$$\equ(4dim)$$

\def\ha{\hskip0.15truein}
\def\ch#1{\centerline{\hbox{#1}}}
\def\vs#1{\vskip#1pt}
\bigskip
{\baselineskip=0pt
\ch{$ \sa1y $} 
\vs{20}
\ch{$ \sc xxx \ha \sc yyx \ha \sc yxy \ha \sc xyy $}
\vs{15}                                               
\ch{$ \sa3x \hskip1truein \sa3y $}                    
\vs{3}                                                
\ch{$ \sa1x $} 
\vs{15}                                               
\ch{$ \sa2y \hskip1.8truein \sa2x $}                  
}
$$\equ(star)$$
\bigskip

\bye